%% file: main.tex
\documentclass{article}
\usepackage{titlesec}
\usepackage{titling}
\usepackage{graphicx} 
\usepackage{amsmath}
\usepackage[margin=1.0in]{geometry}

\usepackage{float}
\usepackage{lscape}
\usepackage{palatino}
\usepackage{verbatim}
\usepackage{authblk}
\usepackage{hyperref}
\usepackage{soul}
\usepackage[numbers,sort&compress]{natbib}
\usepackage{multibib}
\newcites{inv}{Inventory References}
\usepackage[labelfont=bf]{caption}
\usepackage{threeparttable}
\usepackage{threeparttablex}
\usepackage{booktabs}
\usepackage{array}
\usepackage{longtable}
\usepackage{xcolor}
\usepackage{amssymb}
\usepackage{bm}
\usepackage[bottom]{footmisc}

\definecolor{metgray}{gray}{0.45}
\definecolor{DeclineRed}{HTML}{B84545}
\definecolor{IncreaseGreen}{HTML}{3D7A4A}
\definecolor{MixedBlue}{HTML}{4A6FA5}
\definecolor{StableGray}{HTML}{666666}

\hypersetup{
    colorlinks=true,
    linkcolor=black,
    urlcolor=black,
    citecolor=black
}

\title{Is Innovation Becoming Less Disruptive? An Inventory of the Literature\thanks{We thank the Alfred P. Sloan Foundation (grant to R.J.F.), National Science Foundation (grant nos. 1829168, 1932596, and 2318172 to R.J.F. and no. 1829302 to E.L), and Wellcome Leap Foundation (grants to R.J.F. and M.P.) for financial support of work related to this project. The funders had no role in study design, data collection and analysis, or preparation of the manuscript. Address correspondence to R.J.F. (\href{mailto:rfunk@umn.edu}{rfunk@umn.edu}).}}
\author[3]{Xiangting Wu}
\author[3]{Linhui Wu}
\author[1]{Michael Park}
\author[2]{Erin Leahey}
\author[3]{Russell J. Funk}
\affil[1]{{\small{Organisational Behaviour, INSEAD}}}
\affil[2]{{\small{School of Sociology, University of Arizona}}}
\affil[3]{{\small{Carlson School of Management, University of Minnesota}}}
\date{\today}

\begin{document}
\maketitle

\begin{abstract}
A growing literature has examined whether innovation is becoming less disruptive, spanning diverse domains and data sources and using a range of methodologies. This paper provides an inventory of 105 studies exploring this question. The evidence is largely consistent in direction. Studies spanning scientific papers, patents, products, legal cases, music, and visual art consistently report evidence of a decline. This pattern holds not only for citation-based measures, but also for text-based approaches, firm displacement rates, product similarity networks, and audio and visual embeddings. The literature has also identified notable exceptions, including rebounds in specific domains and predictable variation across field lifecycles. We catalog each study's data, methods, and findings to provide a resource for researchers and policymakers seeking to understand the current state of the evidence.
\end{abstract}

\pagebreak

\section{Introduction}

Innovation is fundamental to economic growth, human health, and the progress of society \cite{romer1986increasing, romer1990endogenous}. Yet there is growing concern that innovation may be slowing across many domains of creative activity. Researchers working from diverse disciplines have documented stagnating productivity growth \cite{gordon2017rise, cowen2011great}, the rising burden of knowledge required to reach the frontier \cite{jones2009burden, bloom2020ideas}, declining business dynamism \cite{akcigit2023business, decker2018changing}, and shifts in the organization of science and creative work that may discourage exploration \cite{bhattacharya2020stagnation, azoulay2011incentives}. These patterns have been observed not only in traditional domains of technological and scientific innovation, but also in artistic and humanistic fields---from the visual arts to music---and even in the evolution of legal precedent.

In 2023, Park, Leahey, and Funk contributed to this literature by documenting large-scale evidence in \emph{Nature} that papers and patents have become less disruptive over time. Disruptive work breaks from the past, rendering prior approaches obsolete, and is often contrasted with consolidating work, which builds on and extends existing foundations. Both are essential to scientific and technological advance, but a healthy ecosystem likely requires balance. The finding that science and technology are increasingly consolidating rather than disrupting therefore sparked considerable interest and debate about the nature of scientific and technological progress.

Since then, the literature on declining disruptiveness has continued to grow. Researchers have examined the pattern across diverse domains (from physics to pharmacy, from patents to products), using a variety of different data (Web of Science, USPTO, firm-level databases, audio features, image embeddings), and employing multiple methodological approaches (citation-based metrics, text analysis, product similarity networks).

This paper provides a systematic inventory of 105 studies that bear on the question---Is innovation becoming less disruptive? Our goal is to examine whether independent investigations, using different data and methods, reach similar conclusions.

\section{Identification of Studies}

We identified studies by collecting all citations to four seed articles: Funk and Owen-Smith \cite{funk2017dynamic}, Wu, Wang, and Evans \cite{wu2019large}, Park, Leahey, and Funk \cite{park2023papers}, and Lin, Frey, and Wu \cite{lin2023remote}. These papers were selected because they developed the core methodology for measuring disruptiveness in citation networks and were published in high-visibility venues (\emph{Nature} and \emph{Management Science}), making subsequent empirical work on this topic likely to cite them. Using Dimensions.ai, we retrieved 1,587 citing works. We supplemented these with manual searches to ensure comprehensiveness, though we acknowledge that this citation-based approach may miss relevant work that does not cite these particular papers. As such, our inventory should be considered a conservative lower bound on the total body of evidence.

We included peer-reviewed articles, preprints, book chapters, conference proceedings, and theses, provided they reported original empirical findings. We excluded purely theoretical works, review articles, and papers presenting only computational or formal mathematical models without substantive empirical analysis. We note that the broader literature on disruptiveness is considerably larger than what we catalog here, encompassing work on social and organizational mechanisms that generate disruptive outcomes, methodological refinements to measurement, and validation studies, among other topics. Our inventory focuses exclusively on studies containing empirical information on temporal trends; work without such information falls outside our scope regardless of its other contributions. We included studies regardless of whether they reported decline, stability, or increase in disruptiveness over time.

At the same time, we were deliberate in maintaining a narrow scope. We included only studies that used established disruption indicators (e.g., CD, D, DI) or that explicitly examined temporal trends in ``disruption'' or ``disruptiveness.'' We did \emph{not} include the vast literatures on related constructs---novelty, originality, breakthrough innovation, business dynamism, or economic growth---even though many of these document similar declines. In this way, our sample of 105 reflects a conservative, focused scope.\footnote{We included two studies that use the term ``creative destruction'' rather than ``disruption,'' as our assessment is that these terms were being used synonymously to describe the same underlying phenomenon.}

Our analytic sample spans papers published between 2017 and 2026. We included not only studies where temporal trends were a primary research question, but also those reporting trends as secondary analyses in papers primarily addressing other questions—team composition, novelty, field dynamics, or scientists' careers. The inclusion of such studies helps mitigate concerns about publication bias.

For each study, we summarize the authors' reported findings as closely as possible. In some cases, we add brief interpretive annotations---for instance, noting implications for methodological debates or contextualizing findings relative to other studies. These annotations reflect our assessment and are not claims made by the original authors.

Table~\ref{tab:inventory} presents the full inventory of 105 studies. Each study is assigned an entry number, with the table sorted in reverse chronological order (most recent first). The rightmost column indicates the reported trend---a downward arrow denotes declining disruptiveness, an upward arrow denotes increasing disruptiveness, a horizontal arrow denotes stable or null findings, and mixed arrows denote studies reporting both decline and rebound. Throughout the text, we refer to studies by their entry numbers (e.g., entry~\ref{inv:park2023n}).

\section{Preliminary Synthesis}

Much has been learned about the disruptiveness of innovative activity in recent years. Over 100 studies now bear on the question of whether disruptiveness has declined, with the large majority reporting evidence that it has. This represents a relatively high degree of convergence for a stream of literature that has mostly emerged within the last decade. At the same time, the field continues to grow, with new studies identifying important boundary conditions, making methodological refinements and improvements, addressing shortcomings of earlier work, and isolating potential mechanisms. This section provides a preliminary synthesis of what has been learned.

Large-scale studies examining science and technology as a whole consistently report evidence of decreasing disruptiveness over time (e.g., entries~\ref{inv:wu2019n}, \ref{inv:lin2023n}, \ref{inv:sixt2024qss}, \ref{inv:yu2025wp}), a pattern that holds across decades of data and diverse creative domains. This finding has appeared in journals including \emph{Nature}, \emph{PNAS}, \emph{Nature Human Behaviour}, \emph{Management Science}, \emph{Nature Computational Science}, and \emph{American Sociological Review}, among many others (entries~\ref{inv:park2023n}, \ref{inv:li2024pnas}, \ref{inv:kedrick2024nhb}, \ref{inv:funk2017ms}, \ref{inv:deng2025wp}, \ref{inv:leahey2023asr}). Evidence of decline has been reported not only in scientific papers and patents, but also in legal cases (entry~\ref{inv:lee2024wp}), software, consumer products (entries~\ref{inv:he2024bwp}, \ref{inv:jeong2025s}), music (entry~\ref{inv:falcao2020pismirc}), and visual art (entry~\ref{inv:shinichi2025wp}). Field-specific case studies in medicine---spanning radiology, plastic surgery, craniofacial surgery, pediatric surgery, colorectal surgery, breast cancer, and pharmacy---document similar patterns (entries~\ref{inv:abu-omar2022cr}, \ref{inv:hansdorfer2021prsgo}, \ref{inv:horen2021jcs}, \ref{inv:sullivan2021jps}, \ref{inv:becerra2022dcr}, \ref{inv:grunvald2021apjcp}, \ref{inv:becerra2024ajhp}).

The pattern does not seem to depend on data source. Evidence of declining disruptiveness has been reported across large, open-access databases, including Crossref (entry~\ref{inv:spinellis2023pone}), Microsoft Academic Graph (entries~\ref{inv:zeng2023wp}, \ref{inv:chen2022wp}, \ref{inv:coles2024wp}, \ref{inv:li2024wp}, \ref{inv:naude2025pone}, \ref{inv:risha2024wp}, \ref{inv:shu2023wp}, \ref{inv:tang2024ic}, \ref{inv:tonde2025wp}, \ref{inv:wu2023joi}, \ref{inv:yang2025ipm}, \ref{inv:yang2025jis}, \ref{inv:yang2024joi}, \ref{inv:yang2024aqss}, \ref{inv:yang2024bqss}, \ref{inv:yang2025jdis}, \ref{inv:wu2024s}, \ref{inv:vaughn2025jps}, \ref{inv:lin2023n}, \ref{inv:cui2022wp}), OpenAlex (entries~\ref{inv:jurowetzki2025as}, \ref{inv:lee2023wp}, \ref{inv:li2025wp}, \ref{inv:li2025awp}, \ref{inv:zhang2025s}, \ref{inv:cao2024ic}, \ref{inv:cheng2025wp}, \ref{inv:fang2025wp}, \ref{inv:li2025re}), SciSciNet (entries~\ref{inv:deng2025wp}, \ref{inv:li2024pnas}, \ref{inv:holst2024wp}, \ref{inv:park2025wp}, \ref{inv:yang2024s}, \ref{inv:zheng2025wp}, \ref{inv:zhang2024wp}), and PubMed (entries~\ref{inv:sheng2023s}, \ref{inv:wang2023wp}). The pattern also holds in commercial databases like Dimensions.ai (entries~\ref{inv:sixt2024qss}, \ref{inv:stack2025wp}), Web of Science, and Scopus (entries~\ref{inv:tang2024joi}, \ref{inv:you2022joi}) that often have higher metadata quality. The finding further holds in field-specific databases, where metadata quality is also often higher given the smaller scale and greater opportunity for manual curation (e.g., entries~\ref{inv:chen2025chi}). Many studies have also been done using both U.S. and global patent data (e.g., entries~\ref{inv:zhou2024wp}, \ref{inv:yang2025rp}, \ref{inv:wang2025hssc}, \ref{inv:wang2024eist}, \ref{inv:qu2024rdm}, \ref{inv:macher2024rp}, \ref{inv:lyu2025wp}, \ref{inv:kaltenberg2023rp}, \ref{inv:he2024awp}, \ref{inv:funk2017ms}, \ref{inv:cao2025mm}, \ref{inv:bowen2019wp}, \ref{inv:boot2025jar}).

The pattern also does not seem to be contingent on the specific metric selected. The literature has produced dozens of indicators intended to capture disruptiveness, including CD, DI, D, and numerous variants (entry~\ref{inv:gebhart2023wp}), as well as novel approaches such as the Product Disruption Index (entries~\ref{inv:he2024bwp},  \ref{inv:jeong2025s}), the Journal Disruption Index (entry~\ref{inv:jurowetzki2025as}), decomposed measures of disruption of and by prior work (entry~\ref{inv:deng2025wp}), concept-level disruptiveness based on MeSH terms (entry~\ref{inv:copara2024wp}), the Global Disruption Index (entry~\ref{inv:yang2023issip}), the Disruptive Knowledge Content index (entry~\ref{inv:tang2024joi}), and the Foundation Index (entry~\ref{inv:fang2025wp}). Each metric has specific strengths and weaknesses, and each was developed with particular improvements in mind. Yet evidence of decline appears across approaches. This convergence across operationalizations suggests that the pattern reflects a substantive phenomenon rather than an artifact of a particular measurement approach.

Critically, the finding does not seem to depend on the use of bibliometrics or citation data. While bibliometric data are valuable for studying innovation, they have known limitations, and scholars have appropriately investigated whether findings on declining disruption could be influenced by them. A number of studies have addressed this concern using alternative approaches. Bowen et al.\ (entry~\ref{inv:bowen2019wp}) document declining disruptiveness in U.S.\ patents using text analysis, finding that average disruptive potential fell to roughly one quarter of its peak by 2010. Boot and Vladimirov (entry~\ref{inv:boot2025jar}) confirm the pattern using multiple text-based proxies, as do Huang and Li (entry~\ref{inv:huang2024wp}) for MOEA/D research. Beyond text, Bessen et al.\ (entry~\ref{inv:bessen2020wp}) and Peters and Walsh (entry~\ref{inv:peters2022wp}) show declining disruption using firm displacement rates and other economic indicators derived from Census and Compustat data. He and Lee (entry~\ref{inv:he2024bwp}) and Jeong and Lee (entry~\ref{inv:jeong2025s}) develop product disruption indices based on similarity networks for automobiles and mobile phones, respectively, avoiding bibliometric data entirely. Falcão et al.\ (entry~\ref{inv:falcao2020pismirc}) measure disruption from raw audio features in Brazilian Forró music, and Shinichi and Matsui (entries~\ref{inv:shinichi2024wp} and~\ref{inv:shinichi2025wp}) apply a deep learning adaptation of the disruption index to Japanese woodblock prints. All report evidence of declines. These studies are unlikely to be vulnerable to limitations of citation-based measures.

While the evidence for decline is largely consistent, the literature has also identified notable exceptions. Substantively, these exceptions represent potential bright spots, perhaps offering clues into underlying mechanisms or policy interventions for accelerating innovation. Methodologically, they are also noteworthy---mechanical artifacts would not be expected to produce structured variation. Studies have found rebounds in particular areas or even increasing or stable disruption in specific fields. For example, Vaughn et al.\ (entry~\ref{inv:vaughn2025jps}) document rising disruption in fetal surgery, Tang et al.\ (entry~\ref{inv:tang2024ic}) find an upward trend in AI research after 2014, Macher et al.\ (entry~\ref{inv:macher2024rp}) observe a rebound in highly disruptive patents after 2008 (especially in IT), Wu et al.\ (entry~\ref{inv:wu2023joi}) show that ML-related economics papers became more disruptive after 2010, and Petersen et al.\ (entry~\ref{inv:petersen2025joi}) find stable trends after controls in a 1995–2015 sample (see also entries~\ref{inv:bentley2023acs}, \ref{inv:yu2024s}, \ref{inv:zutphen2023wp}).

The literature is increasingly turning from documentation to explanation. A number of studies have begun to investigate the mechanisms underlying declining disruptiveness, examining factors such as team size (entries~\ref{inv:wu2019n}, \ref{inv:coles2024wp}), career age and workforce composition (entries~\ref{inv:cui2022wp}, \ref{inv:kaltenberg2023rp}, \ref{inv:li2024pnas}), the burden of knowledge (entry~\ref{inv:lyu2025wp}), reliance on narrow slices of prior work (entries~\ref{inv:park2023n}, \ref{inv:yu2025wp}), the growing dominance of established ideas and elite researchers (entries~\ref{inv:liu2025ipm}, \ref{inv:chu2021pnas}), declining topic switching (entry~\ref{inv:yang2025ipm}), reduced conceptual churn (entry~\ref{inv:kedrick2024nhb}), the suppression of paradigm-deviating work through peer review (entry~\ref{inv:shu2023wp}), the rise of remote collaboration (entry~\ref{inv:lin2023n}), the decline of generalist scientists (entry~\ref{inv:risha2024wp}), and firms consciously preventing disruptive innovation (entry~\ref{inv:brauer2024wp}). While no single mechanism has emerged as definitive, the accumulating evidence points toward a constellation of interrelated factors. Understanding these mechanisms matters for policy---whether the goal is to encourage more disruptive work, preserve space for consolidation, or simply ensure a balance.

\section{Discussion}

The 105 studies cataloged here represent a notable convergence of evidence across research teams, data sources, and methodological approaches. The pattern appears in scientific papers, patents, products, legal cases, music, and visual art, and has been documented using not only citation-based measures but also firm displacement rates, patent text, audio features, visual embeddings, and product similarity networks. The literature has also identified plausible exceptions, often in young or rapidly evolving fields. The field is now turning from documentation to explanation, which remains an open and active area of inquiry. This inventory is intended as a resource for researchers and policymakers seeking to understand the current state of the evidence. Across measures and domains, that evidence points consistently toward decline.

\pagebreak

\section{Inventory of Studies}

\input{raw/inventory_table}

\pagebreak

\bibliographystyle{unsrtnat}
\bibliography{references_main}

\pagebreak

\renewcommand{\refname}{Inventory References}
\bibliographystyleinv{unsrtnat}
\bibliographyinv{references_inventory}
\end{document}

%% file: raw/inventory_table.tex
\newcounter{inventry}
\setcounter{inventry}{106}

\begin{longtable}{@{}r p{15cm} c@{}}
\caption{Summary of Findings on Trends in Disruptiveness}\label{tab:inventory} \\
\toprule
\addlinespace[0.5em]
\endfirsthead
\endhead
\bottomrule
\endlastfoot
\multicolumn{3}{c}{\large\textsc{2026} (2 studies)} \\[0.8em]
\setcounter{inventry}{104}\refstepcounter{inventry}\label{inv:cao2025mm}\textcolor{metgray}{105} & \nociteinv{cao2025mm}\textbf{\href{https://doi.org/10.1007/s11023-025-09755-8}{\color{black}Cao \& Cai (2026)}}, \emph{\emph{Minds Mach}} \newline
{\small\color{metgray} AI patents, 1976--2020. $\cdot$ Data: USPTO} \newline
``A noticeable decline over the past two decades\ldots across nearly all AI subcategories'' (Fig. 2). AI ``entered the track of consolidation.'' & {\color{DeclineRed}\Large$\bm{\downarrow}$} \\
\addlinespace[1.2em]

\setcounter{inventry}{103}\refstepcounter{inventry}\label{inv:liu2025ipm}\textcolor{metgray}{104} & \nociteinv{liu2025ipm}\textbf{\href{https://doi.org/10.1016/j.ipm.2025.104382}{\color{black}Liu et al. (2026)}}, \emph{\emph{Inf Process Manag}} \newline
{\small\color{metgray} 647K papers, 1893--2020. $\cdot$ Data: APS} \newline
``Overall disruptiveness\ldots exhibited a consistent decline starting from the 1940s'' (Fig. 5A). Newcomers more disruptive (Fig. 5B) but science increasingly reliant on established elites (Fig. 5C).  & {\color{DeclineRed}\Large$\bm{\downarrow}$} \\
\addlinespace[1.2em]

\\[1.5em]
\multicolumn{3}{c}{\large\textsc{2025} (32 studies)} \\[0.8em]
\setcounter{inventry}{102}\refstepcounter{inventry}\label{inv:becerra2024ajhp}\textcolor{metgray}{103} & \nociteinv{becerra2024ajhp}\textbf{\href{https://doi.org/10.1093/ajhp/zxae309}{\color{black}Becerra et al. (2025)}}, \emph{\emph{Am J Health-Syst Pharm}} \newline
{\small\color{metgray} Pharmacy, 1954--2014. $\cdot$ Data: PubMed} \newline
``Despite the many disruptive papers, there was a lack of more recent papers, indicating the need for modern innovation in pharmacy'' (Fig. 4).  & {\color{DeclineRed}\Large$\bm{\downarrow}$} \\
\addlinespace[1.2em]

\setcounter{inventry}{101}\refstepcounter{inventry}\label{inv:boot2025jar}\textcolor{metgray}{102} & \nociteinv{boot2025jar}\textbf{\href{https://doi.org/10.1111/1475-679x.12580}{\color{black}Boot \& Vladimirov (2025)}}, \emph{\emph{J Account Res}} \newline
{\small\color{metgray} U.S. patents, AIPA (2000) policy shock. $\cdot$ Data: USPTO} \newline
Exploiting AIPA for causal inference, reports ``the nature of patented innovation has changed, with the proportion of nondisruptive patents increasing substantially.'' Multiple disruption proxies---citation and text-based---show decline (Fig. 5). & {\color{DeclineRed}\Large$\bm{\downarrow}$} \\
\addlinespace[1.2em]

\setcounter{inventry}{100}\refstepcounter{inventry}\label{inv:chen2025chi}\textcolor{metgray}{101} & \nociteinv{chen2025chi}\textbf{\href{https://doi.org/10.1145/3706598.3713917}{\color{black}Chen \& Li (2025)}}, \emph{CHI} \newline
{\small\color{metgray} HCI, 1982--2023. $\cdot$ Data: ACM DL, OpenAlex} \newline
``The growth rate of disruptive works does not match up with that of the explosion of published papers'' (Figs. 2, 3). Robustness checks (Appendix C, D) further validate decline. & {\color{DeclineRed}\Large$\bm{\downarrow}$} \\
\addlinespace[1.2em]

\setcounter{inventry}{99}\refstepcounter{inventry}\label{inv:cheng2025wp}\textcolor{metgray}{100} & \nociteinv{cheng2025wp}\textbf{\href{https://doi.org/10.51408/issi2025_009}{\color{black}Cheng \& Ma (2025)}}, \emph{ISSI} \newline
{\small\color{metgray} Scientometrics, 1955--2024. $\cdot$ Data: OpenAlex, WoS} \newline
Using Main Path Analysis, concludes ``the field of scientometrics is experiencing a decline in disruption'' (Figs. 7--8).  & {\color{DeclineRed}\Large$\bm{\downarrow}$} \\
\addlinespace[1.2em]

\setcounter{inventry}{98}\refstepcounter{inventry}\label{inv:cheng2025s}\textcolor{metgray}{99} & \nociteinv{cheng2025s}\textbf{\href{https://doi.org/10.1007/s11192-025-05285-3}{\color{black}Cheng et al. (2025)}}, \emph{\emph{Scientometrics}} \newline
{\small\color{metgray} Earth Sciences, 2000--2019. $\cdot$ Data: WoS} \newline
Statistical analysis reveals declining disruption over time; Disruption$\sim$Year: r=-0.227, p$<$0.001 (Table 2). & {\color{DeclineRed}\Large$\bm{\downarrow}$} \\
\addlinespace[1.2em]

\setcounter{inventry}{97}\refstepcounter{inventry}\label{inv:deng2025wp}\textcolor{metgray}{98} & \nociteinv{deng2025wp}\textbf{\href{https://doi.org/10.1038/s43588-025-00808-7}{\color{black}Deng et al. (2025)}}, \emph{\emph{Nat Comput Sci}} \newline
{\small\color{metgray} 1900--2020, 90M papers. $\cdot$ Data: APS, SciSciNet} \newline
Confirms that ``both $d^r$ [disruption of prior work] and $d^c$ [disruption by future work] decline with publication year'' (Fig. S2c), attributing this to ``the growing prevalence of consolidatory work in modern science.'' While the fraction of ``persistently disruptive'' papers is rising (Fig. 5c), this is driven by the collapse of $d^c$; papers today are ``more likely to be consolidated than to be disrupted,'' even as their own ability to disrupt the past ($d^r$) steadily fades (Figs. 5d, S10b). & {\color{DeclineRed}\Large$\bm{\downarrow}$} \\
\addlinespace[1.2em]

\setcounter{inventry}{96}\refstepcounter{inventry}\label{inv:fang2025wp}\textcolor{metgray}{97} & \nociteinv{fang2025wp}\textbf{\href{https://arxiv.org/abs/2510.03240v2}{\color{black}Fang \& Evans (2025)}}, \emph{arXiv} \newline
{\small\color{metgray} 1950--2023, 23M/19M papers. $\cdot$ Data: OpenAlex, WoS} \newline
Provides evidence for declining disruptiveness by showing ``foundational\ldots work within fields has declined in recent years'' (Fig. 4). The ``Foundation Index'' captures work that ``eclipses prior ideas'' and is thus ``closely aligned with the original Disruption Index.'' They further note declining ``generalization'' (cross-field combinations), suggesting a shift toward synthesizing rather than disrupting existing ideas. & {\color{DeclineRed}\Large$\bm{\downarrow}$} \\
\addlinespace[1.2em]

\setcounter{inventry}{95}\refstepcounter{inventry}\label{inv:he2024bwp}\textcolor{metgray}{96} & \nociteinv{he2024bwp}\textbf{\href{https://doi.org/10.1016/j.joi.2025.101703}{\color{black}He \& Lee (2025)}}, \emph{\emph{J Informetr}} \newline
{\small\color{metgray} Automobiles, 2014--2023, 4,114 car models. $\cdot$ Data: Edmunds} \newline
Introduces a Product Disruption Index based on product similarity networks, avoiding bibliometric data entirely. Reports ``a clear downward trend\ldots mirroring the overarching trend of diminishing innovation disruptiveness over time'' (Fig. 4b). ``Earlier [car] models\ldots are more likely to be disruptive.'' & {\color{DeclineRed}\Large$\bm{\downarrow}$} \\
\addlinespace[1.2em]

\setcounter{inventry}{94}\refstepcounter{inventry}\label{inv:jeong2025s}\textcolor{metgray}{95} & \nociteinv{jeong2025s}\textbf{\href{https://doi.org/10.1007/s11192-025-05395-y}{\color{black}Jeong \& Lee (2025)}}, \emph{\emph{Scientometrics}} \newline
{\small\color{metgray} 1994--2023, 12,554 mobile products. $\cdot$ Data: GSMArena} \newline
Applies CD index on product similarity network (shared OS, camera, chipset). ``Products with the highest disruption index for all four companies are their initial product lines'' (Fig. 10). ``Most firms introduce more disruptive products in early stages\ldots followed by gradual decline.'' & {\color{DeclineRed}\Large$\bm{\downarrow}$} \\
\addlinespace[1.2em]

\setcounter{inventry}{93}\refstepcounter{inventry}\label{inv:jiang2025jsp}\textcolor{metgray}{94} & \nociteinv{jiang2025jsp}\textbf{\href{https://doi.org/10.3138/jsp-2023-0056}{\color{black}Jiang et al. (2025)}}, \emph{\emph{J Sch Publ}} \newline
{\small\color{metgray} Asian journals, 2010--2019. $\cdot$ Data: OpenCitations} \newline
Uses a Journal Disruption Index (JDI) to evaluate journals from China, India, S. Korea, finding an upward trend 2010--2019. But the authors also note ``disruptive innovation level\ldots still far from Nature and Science,'' implying catching up from lower baseline.  & {\color{IncreaseGreen}\Large$\bm{\uparrow}$} \\
\addlinespace[1.2em]

\setcounter{inventry}{92}\refstepcounter{inventry}\label{inv:jurowetzki2025as}\textcolor{metgray}{93} & \nociteinv{jurowetzki2025as}\textbf{\href{https://doi.org/10.1007/s00146-024-02171-z}{\color{black}Jurowetzki et al. (2025)}}, \emph{\emph{AI Soc}} \newline
{\small\color{metgray} AI research, 2000--2021, 1.7M papers, 143K researchers. $\cdot$ Data: OpenAlex} \newline
Finds academia-to-industry transitions are rising (Fig. 2c), and that post-transition, researchers show ``reduced disruptive influence'' (Table 2). ``As researchers spend more time in industry, they produce less cited, disruptive, and novel science.'' & {\color{DeclineRed}\Large$\bm{\downarrow}$} \\
\addlinespace[1.2em]

\setcounter{inventry}{91}\refstepcounter{inventry}\label{inv:li2025wp}\textcolor{metgray}{92} & \nociteinv{li2025wp}\textbf{\href{https://arxiv.org/abs/2506.15959v4}{\color{black}Li et al. (2025)}}, \emph{Working Paper} \newline
{\small\color{metgray} 1965--2020, 41M papers. $\cdot$ Data: OpenAlex} \newline
Documents ``the decline of disruption in science over time'' (Sec. 4.1). Proportion of highly disruptive papers (top 5\%) fell from $\sim$50\% to below 20\% (Fig. 2). & {\color{DeclineRed}\Large$\bm{\downarrow}$} \\
\addlinespace[1.2em]

\setcounter{inventry}{90}\refstepcounter{inventry}\label{inv:li2025awp}\textcolor{metgray}{91} & \nociteinv{li2025awp}\textbf{\href{https://arxiv.org/abs/2512.03723v1}{\color{black}Li et al. (2025)}}, \emph{arXiv} \newline
{\small\color{metgray} 1965--2020, 41M papers. $\cdot$ Data: OpenAlex} \newline
Finds the ``share of disruptive papers (D $>$ 0) declines from nearly 50\% in 1965 to under 20\% by 2020'' (Figs. 14--15). However, ``absolute number of highly disruptive papers has remained stable over the past six decades\ldots consistent with the 'conservation of highly disruptive work' (Park et al., 2023).'' (Fig. 7) & {\color{DeclineRed}\Large$\bm{\downarrow}$} \\
\addlinespace[1.2em]

\setcounter{inventry}{89}\refstepcounter{inventry}\label{inv:li2025re}\textcolor{metgray}{90} & \nociteinv{li2025re}\textbf{\href{https://doi.org/10.1093/reseval/rvaf026}{\color{black}Li et al. (2025)}}, \emph{\emph{Res Eval}} \newline
{\small\color{metgray} Social sciences, 1997--2017, 1.09M papers. $\cdot$ Data: OpenAlex, WoS} \newline
``Figure 2a presents the declining trend of the CD5 over the last 20 years. This downtrend is similar across disciplines.'' The authors further report after 2007 ``the growth rate of disruptive innovative papers was flat'' while consolidating papers grew rapidly (Fig. 2c). Validated with two text-based measures of disruption, showing results not driven by citation-data artifacts.  & {\color{DeclineRed}\Large$\bm{\downarrow}$} \\
\addlinespace[1.2em]

\setcounter{inventry}{88}\refstepcounter{inventry}\label{inv:lyu2025wp}\textcolor{metgray}{89} & \nociteinv{lyu2025wp}\textbf{Lyu (2025)}, \emph{Working Paper} \newline
{\small\color{metgray} U.S. patents, 1977--2006, 308K patents. $\cdot$ Data: USPTO} \newline
Shows steady decline in disruptiveness (Fig. 2). ``Overall impact of knowledge accumulation on innovation is found to be negative.'' Results supported by regression analysis and formal model (Table 4).  & {\color{DeclineRed}\Large$\bm{\downarrow}$} \\
\addlinespace[1.2em]

\setcounter{inventry}{87}\refstepcounter{inventry}\label{inv:naude2025pone}\textcolor{metgray}{88} & \nociteinv{naude2025pone}\textbf{\href{https://doi.org/10.1371/journal.pone.0323297}{\color{black}Naud{\'e} (2025)}}, \emph{\emph{PLoS ONE}} \newline
{\small\color{metgray} Business and Entrepreneurship, 1961--2020, 68K papers. $\cdot$ Data: MAG} \newline
Reports ``the average disruption score of papers have declined by a factor of 36 between the 1960s and the 2010s'' (Fig. 2), and ``the share of papers with a D$>$0 has consistently declined over time'' (Fig. 3).  & {\color{DeclineRed}\Large$\bm{\downarrow}$} \\
\addlinespace[1.2em]

\setcounter{inventry}{86}\refstepcounter{inventry}\label{inv:park2025wp}\textcolor{metgray}{87} & \nociteinv{park2025wp}\textbf{\href{https://arxiv.org/abs/2503.00184v1}{\color{black}Park et al. (2025)}}, \emph{arXiv} \newline
{\small\color{metgray} 1945--2010 (papers, 22M), 1980--2010 (patents, 2.9M). $\cdot$ Data: SciSciNet, USPTO, WoS} \newline
Responds to Holst et al.'s critique that declining disruptiveness is an artifact of zero-backward-citation papers. Using critics' own dataset, metric, and regression model, finds ``statistically and practically significant declines\ldots that equal major benchmark transformations in science'' (Figs. 1--2, p$<$0.001). Decline magnitude equals the gap between an average paper and a Nobel winner. Critics' own regression (their Table S1) shows significant decline---``a finding they neither acknowledge nor interpret.'' & {\color{DeclineRed}\Large$\bm{\downarrow}$} \\
\addlinespace[1.2em]

\setcounter{inventry}{85}\refstepcounter{inventry}\label{inv:petersen2025joi}\textcolor{metgray}{86} & \nociteinv{petersen2025joi}\textbf{\href{https://doi.org/10.1016/j.joi.2024.101605}{\color{black}Petersen et al. (2025)}}, \emph{\emph{J Informetr}} \newline
{\small\color{metgray} 1995--2015, 7.8M papers. $\cdot$ Data: SciSciNet} \newline
Argues that after controls, trends are ``at the level of noise.'' However, the study covers 1995--2015, a period that falls after the steepest decline documented in larger-scale studies spanning 1945--1990. Additionally, their preprocessing excludes papers with fewer than 10 references and more than 1,000 citations---restrictions that may bias against finding disruptive work. & {\color{StableGray}\Large$\bm{\rightarrow}$} \\
\addlinespace[1.2em]

\setcounter{inventry}{84}\refstepcounter{inventry}\label{inv:qu2024rdm}\textcolor{metgray}{85} & \nociteinv{qu2024rdm}\textbf{\href{https://doi.org/10.1111/radm.12685}{\color{black}Qu et al. (2025)}}, \emph{\emph{R\&D Manage}} \newline
{\small\color{metgray} Electric communication industry, 1995--2019, 29.6K patents. $\cdot$ Data: USPTO} \newline
Documents a significant negative effect of year on disruption ($\beta$ = -0.0592, p$<$0.01, Table 2). & {\color{DeclineRed}\Large$\bm{\downarrow}$} \\
\addlinespace[1.2em]

\setcounter{inventry}{83}\refstepcounter{inventry}\label{inv:shinichi2025wp}\textcolor{metgray}{84} & \nociteinv{shinichi2025wp}\textbf{\href{https://arxiv.org/abs/2505.08284v1}{\color{black}Shinichi \& Matsui (2025)}}, \emph{arXiv} \newline
{\small\color{metgray} Visual Art (Ukiyo-e), 11,000 images. $\cdot$ Data: Ukiyo-e images} \newline
Using a novel deep learning adaptation of the Disruption Index (DI) to quantify the disruptiveness of images, the authors find ``the disruptiveness of works diminishes over the data period, consistent with the dynamics\ldots of scientific works'' (Fig. 8a). However, within specific schools ``disruptiveness\ldots remained stable'' (Fig. 8b). This divergence, combined with the use of non-citation data, suggests that downward trends in DI reported for scientific domains are not driven by mechanical artifacts. & {\color{DeclineRed}\Large$\bm{\downarrow}$} \\
\addlinespace[1.2em]

\setcounter{inventry}{82}\refstepcounter{inventry}\label{inv:stack2025wp}\textcolor{metgray}{83} & \nociteinv{stack2025wp}\textbf{Stack (2025)}, \emph{Working Paper} \newline
{\small\color{metgray} 1900--2000, all fields. $\cdot$ Data: Dimensions} \newline
Reports ``disruptiveness as measured by the CD index across all research fields has been falling since the 19th century'' (Fig. 2.2). & {\color{DeclineRed}\Large$\bm{\downarrow}$} \\
\addlinespace[1.2em]

\setcounter{inventry}{81}\refstepcounter{inventry}\label{inv:tonde2025wp}\textcolor{metgray}{82} & \nociteinv{tonde2025wp}\textbf{Tonde (2025)}, \emph{Thesis} \newline
{\small\color{metgray} 1960--2020, 52.9M papers. $\cdot$ Data: MAG} \newline
``Our results indicate a steady decline in CD index scores, suggesting that more recent publications tend to reinforce existing knowledge frameworks rather than introduce transformative innovations.'' (Figs. 3.14--3.19). & {\color{DeclineRed}\Large$\bm{\downarrow}$} \\
\addlinespace[1.2em]

\setcounter{inventry}{80}\refstepcounter{inventry}\label{inv:vaughn2025jps}\textcolor{metgray}{81} & \nociteinv{vaughn2025jps}\textbf{\href{https://doi.org/10.1016/j.jpedsurg.2025.162338}{\color{black}Vaughn et al. (2025)}}, \emph{\emph{J Pediatr Surg}} \newline
{\small\color{metgray} Fetal surgery, 1975--2021. $\cdot$ Data: MAG, PubMed} \newline
``While disruption across scientific fields has been on the decline since the turn of the millennium\ldots fetal surgery tells a different story. Here, disruption is not only thriving but also increasing over time'' (Fig. 2). This exception---in a young, evolving field---suggests the decline observed elsewhere is not a mechanical artifact, since such artifacts would appear uniformly across domains. & {\color{IncreaseGreen}\Large$\bm{\uparrow}$} \\
\addlinespace[1.2em]

\setcounter{inventry}{79}\refstepcounter{inventry}\label{inv:wang2025hssc}\textcolor{metgray}{80} & \nociteinv{wang2025hssc}\textbf{\href{https://doi.org/10.1057/s41599-025-06038-6}{\color{black}Wang \& Maynard (2025)}}, \emph{\emph{Humanit Soc Sci Commun}} \newline
{\small\color{metgray} U.S. patents, 1976--2021, 3.7M. $\cdot$ Data: USPTO} \newline
Mean CD5 disruption falls steadily toward zero by 2010 across all gender collaboration types, reinforcing evidence of a broad decline in technological disruption (Fig. 3N). & {\color{DeclineRed}\Large$\bm{\downarrow}$} \\
\addlinespace[1.2em]

\setcounter{inventry}{78}\refstepcounter{inventry}\label{inv:yang2025ipm}\textcolor{metgray}{79} & \nociteinv{yang2025ipm}\textbf{\href{https://doi.org/10.1016/j.ipm.2025.104171}{\color{black}Yang (2025)}}, \emph{\emph{Inf Process Manag}} \newline
{\small\color{metgray} 1950--2020, 1.4M scientists, 27.6M papers. $\cdot$ Data: MAG} \newline
Topic switching is a key driver of disruption; a 1SD increase in switching makes papers 39\% more likely to be disruptive (Fig. 6). Yet ``real switch scores decline over time'' (Figs. 3--5). As switching declines, so does disruption. & {\color{DeclineRed}\Large$\bm{\downarrow}$} \\
\addlinespace[1.2em]

\setcounter{inventry}{77}\refstepcounter{inventry}\label{inv:yang2025jis}\textcolor{metgray}{78} & \nociteinv{yang2025jis}\textbf{\href{https://doi.org/10.1177/01655515251330614}{\color{black}Yang (2025)}}, \emph{\emph{J Inf Sci}} \newline
{\small\color{metgray} 1960--2020, 38.9M papers. $\cdot$ Data: MAG} \newline
Disruptive potential ``drops by 62.5\%, from 0.40 to 0.15'' (Fig. 1f), as ``disruption---measured by how much research challenges existing scientific paradigms---has been on the decline.'' & {\color{DeclineRed}\Large$\bm{\downarrow}$} \\
\addlinespace[1.2em]

\setcounter{inventry}{76}\refstepcounter{inventry}\label{inv:yang2025rp}\textcolor{metgray}{77} & \nociteinv{yang2025rp}\textbf{\href{https://doi.org/10.1016/j.respol.2025.105295}{\color{black}Yang (2025)}}, \emph{\emph{Res Policy}} \newline
{\small\color{metgray} 1980--2017, 6M+ patents. $\cdot$ Data: USPTO} \newline
``The CD index has experienced a steady decline over time'' and ``average CD5 has shown a clear decline since 1980'' (Fig. 4d). The author compares the citation-based CD index to a text-based ``breakthrough'' measure (KI index), finding that the citation-based decline is steady while the text-based measure fluctuates with economic cycles. & {\color{DeclineRed}\Large$\bm{\downarrow}$} \\
\addlinespace[1.2em]

\setcounter{inventry}{75}\refstepcounter{inventry}\label{inv:yang2025jdis}\textcolor{metgray}{76} & \nociteinv{yang2025jdis}\textbf{\href{https://doi.org/10.2478/jdis-2025-0029}{\color{black}Yang et al. (2025)}}, \emph{\emph{J Data Inf Sci}} \newline
{\small\color{metgray} 1950--2016, 29M articles. $\cdot$ Data: MAG, OpenAlex} \newline
Shows disruptiveness (CD index) declined from 0.09 in 1950 to 0.03 in 2016 (Figure 5c). Addresses the ``disruptiveness-citation puzzle'', i.e., highly disruptive papers are increasingly failing to garner high citation impact. & {\color{DeclineRed}\Large$\bm{\downarrow}$} \\
\addlinespace[1.2em]

\setcounter{inventry}{74}\refstepcounter{inventry}\label{inv:yu2025wp}\textcolor{metgray}{75} & \nociteinv{yu2025wp}\textbf{\href{https://arxiv.org/abs/2504.09589v2}{\color{black}Yu et al. (2025)}}, \emph{arXiv} \newline
{\small\color{metgray} 1950--2024, 114M papers. $\cdot$ Data: OECD Patents, OpenAlex, SciSciNet, WoS} \newline
The authors ``find a systematic decline in disruption over time, consistent with prior findings'' (Figs. 2C, 4B). This conclusion holds across multiple large-scale databases, including Web of Science (Figs. S19C, S21B), OpenAlex (Figs. 2C, 4B), and SciSciNet (Figs. S10C, S12B). The study conducts extensive robustness checks to rule out confounds from  citation inflation (Fig. S1, S7) and zero-reference works. Further finds that knowledge independence has declined substantially over time (Fig. 1C), yet ``knowledge independence strongly predicts disruption'' (Fig. 2, Table 1), providing an additional indirect, mechanism-driven explanation for the decline. & {\color{DeclineRed}\Large$\bm{\downarrow}$} \\
\addlinespace[1.2em]

\setcounter{inventry}{73}\refstepcounter{inventry}\label{inv:zhang2025s}\textcolor{metgray}{74} & \nociteinv{zhang2025s}\textbf{\href{https://doi.org/10.1007/s11192-025-05319-w }{\color{black}Zhang et al. (2025)}}, \emph{\emph{Scientometrics}} \newline
{\small\color{metgray} Synchrotron research, 95,720 articles. $\cdot$ Data: OpenAlex} \newline
In models controlling for team characteristics, publication year is a significant negative predictor of disruption (all p$<$0.001). Consistent across DI3 (Table 3) and DI5 robustness checks (Tables 5--6). & {\color{DeclineRed}\Large$\bm{\downarrow}$} \\
\addlinespace[1.2em]

\setcounter{inventry}{72}\refstepcounter{inventry}\label{inv:zheng2025wp}\textcolor{metgray}{73} & \nociteinv{zheng2025wp}\textbf{\href{https://arxiv.org/abs/2509.06212v1}{\color{black}Zheng \& Hou (2025)}}, \emph{arXiv} \newline
{\small\color{metgray} 1960--2020. $\cdot$ Data: SciSciNet} \newline
Identifies four knowledge-production modes; the elite-driven mode, which ``performed most prominently'' on disruption, declines from 59.0\% of papers in the 1960s to 10.4\% in the 2020s (Fig. 4b; Fig. S4). & {\color{DeclineRed}\Large$\bm{\downarrow}$} \\
\addlinespace[1.2em]

\setcounter{inventry}{71}\refstepcounter{inventry}\label{inv:zhou2024wp}\textcolor{metgray}{72} & \nociteinv{zhou2024wp}\textbf{\href{https://doi.org/10.1016/j.eswa.2025.127355 }{\color{black}Zhou et al. (2025)}}, \emph{\emph{Expert Syst Appl}} \newline
{\small\color{metgray} 1976--2022, 8.3M patents. $\cdot$ Data: USPTO} \newline
The authors conclude that ``in terms of scientific value\ldots patents are becoming less disruptive'', attributed to a ``dilution effect.'' & {\color{DeclineRed}\Large$\bm{\downarrow}$} \\
\addlinespace[1.2em]

\\[1.5em]
\multicolumn{3}{c}{\large\textsc{2024} (34 studies)} \\[0.8em]
\setcounter{inventry}{70}\refstepcounter{inventry}\label{inv:brauer2024wp}\textcolor{metgray}{71} & \nociteinv{brauer2024wp}\textbf{Br{\"a}uer (2024)}, \emph{Working Paper} \newline
{\small\color{metgray} Global patents, 1990--2010. $\cdot$ Data: PATSTAT} \newline
Documents declining disruptiveness (Figs. 2, 3), then develops a calibrated model attributing the decline to firms ``consciously preventing disruptive innovation,'' explaining ``52\% of the decline of disruptive innovation until 2010'' (Fig. 6). & {\color{DeclineRed}\Large$\bm{\downarrow}$} \\
\addlinespace[1.2em]

\setcounter{inventry}{69}\refstepcounter{inventry}\label{inv:cao2024ic}\textcolor{metgray}{70} & \nociteinv{cao2024ic}\textbf{\href{https://doi.org/10.1007/978-3-031-57860-1_23}{\color{black}Cao et al. (2024)}}, \emph{iConference} \newline
{\small\color{metgray} Scientometrics research, 6,562 papers. $\cdot$ Data: OpenAlex, WoS} \newline
Disruptiveness declines across all four major topics, with ``a decreasing trend in the disruptive indicator\ldots suggesting a diminishing generation of novel knowledge'' (Figs. 4b, 7).  & {\color{DeclineRed}\Large$\bm{\downarrow}$} \\
\addlinespace[1.2em]

\setcounter{inventry}{68}\refstepcounter{inventry}\label{inv:coles2024wp}\textcolor{metgray}{69} & \nociteinv{coles2024wp}\textbf{Coles (2024)}, \emph{PsyArXiv} \newline
{\small\color{metgray} 1811--2015, 29M papers. $\cdot$ Data: MAG} \newline
``Big team science efforts tend to build upon---vs. disrupt---pre-existing lines of thinking'' (Fig. 3). Rising prevalence of big teams (Fig. 1) implies compositional shift away from disruption. & {\color{DeclineRed}\Large$\bm{\downarrow}$} \\
\addlinespace[1.2em]

\setcounter{inventry}{67}\refstepcounter{inventry}\label{inv:copara2024wp}\textcolor{metgray}{68} & \nociteinv{copara2024wp}\textbf{\href{https://arxiv.org/abs/2406.18792v4}{\color{black}Copara et al. (2024)}}, \emph{arXiv} \newline
{\small\color{metgray} Biomedicine, 2014--2021, 33M papers. $\cdot$ Data: PubMed} \newline
Develops a novel disruptiveness measure for concepts. Concludes ``disruptiveness, influence, and usefulness decrease over time'' (Fig. 2); holds statistically across all top-level MeSH categories (p$<$0.001). Authors note, ``Park and colleagues also corroborate the decreasing trend of disruptiveness, although their analysis was based on scientific articles, in contrast to ours, which is based on MeSH concepts.'' & {\color{DeclineRed}\Large$\bm{\downarrow}$} \\
\addlinespace[1.2em]

\setcounter{inventry}{66}\refstepcounter{inventry}\label{inv:fernandez2024villaverde}\textcolor{metgray}{67} & \nociteinv{fernandez2024villaverde}\textbf{Fern{\'a}ndez-Villaverde et al. (2024)}, \emph{Working Paper} \newline
{\small\color{metgray} 1970--2019. $\cdot$ Data: Census, Compustat, USPTO} \newline
Provides evidence of declining disruption (``creative destruction'') using economic indicators (e.g., firm entry) rather than bibliometrics (Fig. 1b). The authors explicitly state that ``creative destruction\ldots [has] fallen substantially in the US over the past twenty years'', confirming the trend is visible in market behavior independent of citation data. & {\color{DeclineRed}\Large$\bm{\downarrow}$} \\
\addlinespace[1.2em]

\setcounter{inventry}{65}\refstepcounter{inventry}\label{inv:he2024awp}\textcolor{metgray}{66} & \nociteinv{he2024awp}\textbf{\href{https://arxiv.org/abs/2412.19662v1}{\color{black}He et al. (2024)}}, \emph{arXiv} \newline
{\small\color{metgray} 1980--2010, 3.4M patents. $\cdot$ Data: USPTO} \newline
``The average CD index displays a downward trend over time'' with field-level declines 79.2\%--99.6\% (Fig. 3a). & {\color{DeclineRed}\Large$\bm{\downarrow}$} \\
\addlinespace[1.2em]

\setcounter{inventry}{64}\refstepcounter{inventry}\label{inv:holst2024wp}\textcolor{metgray}{65} & \nociteinv{holst2024wp}\textbf{\href{https://arxiv.org/abs/2402.14583v1}{\color{black}Holst et al. (2024)}}, \emph{arXiv} \newline
{\small\color{metgray} Papers and patents (artifact critique). $\cdot$ Data: DBLP, SciSciNet, USPTO} \newline
Using models explicitly designed to correct alleged biases, the authors still find negative and statistically significant year effects on disruptiveness for papers (p$<$0.01) and patents (p$<$0.01) across all specifications (Table S1). & {\color{DeclineRed}\Large$\bm{\downarrow}$} \\
\addlinespace[1.2em]

\setcounter{inventry}{63}\refstepcounter{inventry}\label{inv:huang2024wp}\textcolor{metgray}{64} & \nociteinv{huang2024wp}\textbf{\href{https://arxiv.org/abs/2404.14228v1}{\color{black}Huang \& Li (2024)}}, \emph{arXiv} \newline
{\small\color{metgray} MOEA/D research, 2009--2022. $\cdot$ Data: WoS} \newline
Using text- and citation-based indicators of disruption, the authors find that ``MOEA/D research is becoming less disruptive over time'' (Fig. 15). & {\color{DeclineRed}\Large$\bm{\downarrow}$} \\
\addlinespace[1.2em]

\setcounter{inventry}{62}\refstepcounter{inventry}\label{inv:kedrick2024nhb}\textcolor{metgray}{63} & \nociteinv{kedrick2024nhb}\textbf{\href{https://doi.org/10.1038/s41562-024-01957-x}{\color{black}Kedrick et al. (2024)}}, \emph{\emph{Nat Hum Behav}} \newline
{\small\color{metgray} Physics and social sciences, 1965--2016, 2.6M papers. $\cdot$ Data: APS, WoS} \newline
Conceptual churn predicts disruption ($\beta$ $\approx$ 0.17--0.33, p$<$0.001). Core concept churn declined 15.96--16.82\% (Fig. 2). ``Increasing rigidity'' suggests ``innovative potential of a field declines over time.'' & {\color{DeclineRed}\Large$\bm{\downarrow}$} \\
\addlinespace[1.2em]

\setcounter{inventry}{61}\refstepcounter{inventry}\label{inv:lee2024wp}\textcolor{metgray}{62} & \nociteinv{lee2024wp}\textbf{\href{https://arxiv.org/abs/2410.04493v1}{\color{black}Lee et al. (2024)}}, \emph{arXiv} \newline
{\small\color{metgray} U.S. case law, 1800--2000, 3.87M U.S. cases. $\cdot$ Data: Caselaw Access Project} \newline
Reports that D fell from 0.807 to 0.003 (Fig. 3). Authors conclude ``the legal system is becoming less disruptive.'' & {\color{DeclineRed}\Large$\bm{\downarrow}$} \\
\addlinespace[1.2em]

\setcounter{inventry}{60}\refstepcounter{inventry}\label{inv:li2024pnas}\textcolor{metgray}{61} & \nociteinv{li2024pnas}\textbf{\href{https://doi.org/10.1073/pnas.2322462121}{\color{black}Li et al. (2024)}}, \emph{\emph{PNAS}} \newline
{\small\color{metgray} Scientific careers, cohorts 1960--1990. $\cdot$ Data: APS, SciSciNet} \newline
Holding productivity constant, later cohorts are systematically less disruptive---1990 entrants underperform 1960 cohorts at the same output levels (Figs. 2C, S1C, Table S1); ``the proportion of new papers that continue existing scientific ideas increases and the proportion of disruptive ideas decreases.'' & {\color{DeclineRed}\Large$\bm{\downarrow}$} \\
\addlinespace[1.2em]

\setcounter{inventry}{59}\refstepcounter{inventry}\label{inv:li2024wp}\textcolor{metgray}{60} & \nociteinv{li2024wp}\textbf{\href{https://doi.org/10.48550/arXiv.2402.16839}{\color{black}Li et al. (2024)}}, \emph{arXiv} \newline
{\small\color{metgray} 1965--2020, 41M papers. $\cdot$ Data: MAG} \newline
Disruptive papers ``declined from 50\% to 25\%'' (Fig. 1a, Fig. S3a). Authors address artifact concerns and conclude the trend ``is unlikely to be a pure artifact'' (S2.6). & {\color{DeclineRed}\Large$\bm{\downarrow}$} \\
\addlinespace[1.2em]

\setcounter{inventry}{58}\refstepcounter{inventry}\label{inv:macher2024rp}\textcolor{metgray}{59} & \nociteinv{macher2024rp}\textbf{\href{https://doi.org/10.1016/j.respol.2024.104992}{\color{black}Macher et al. (2024)}}, \emph{\emph{Res Policy}} \newline
{\small\color{metgray} U.S. patents, 1976--2016, 5.3M patents. $\cdot$ Data: USPTO} \newline
The authors evaluate trends in disruptiveness in patents after adding in pre-1976 references and patent applications. Finds that the decline persists, though more modestly, and a rebound in the number of highly disruptive patents after 2008, especially in IT (Figs. 5--6).  & {\color{DeclineRed}\Large$\bm{\downarrow}$\hspace{-3pt}}{\color{IncreaseGreen}\Large$\bm{\uparrow}$} \\
\addlinespace[1.2em]

\setcounter{inventry}{57}\refstepcounter{inventry}\label{inv:martini2024wp}\textcolor{metgray}{58} & \nociteinv{martini2024wp}\textbf{\href{https://doi.org/10.17863/CAM.111428}{\color{black}Martini et al. (2024)}}, \emph{Working Paper} \newline
{\small\color{metgray} Database software, 1983--1993, 51 patents. $\cdot$ Data: Patbase} \newline
Finds ``the average technological disruptiveness\ldots decreased over the years. The same trend can be found for radicalness and originality.'' Concludes ``the inventive step of the patents appears to decrease continuously'' (Fig. 2). & {\color{DeclineRed}\Large$\bm{\downarrow}$} \\
\addlinespace[1.2em]

\setcounter{inventry}{56}\refstepcounter{inventry}\label{inv:petersen2024qss}\textcolor{metgray}{57} & \nociteinv{petersen2024qss}\textbf{\href{https://doi.org/10.1162/qss_a_00333}{\color{black}Petersen et al. (2024)}}, \emph{\emph{Quant Sci Stud}} \newline
{\small\color{metgray} 1945--2012, 30M papers. $\cdot$ Data: MAG} \newline
Replicates decline in disruptiveness reported by other studies, but attributes it to ``citation inflation.'' Computational model reproduces decline. However, the simulation is not compared to empirical trends, and whether decline persists after the proposed citation inflation adjustments are made remains unclear. & {\color{DeclineRed}\Large$\bm{\downarrow}$} \\
\addlinespace[1.2em]

\setcounter{inventry}{55}\refstepcounter{inventry}\label{inv:risha2024wp}\textcolor{metgray}{56} & \nociteinv{risha2024wp}\textbf{\href{https://arxiv.org/abs/2304.14518v2}{\color{black}Risha et al. (2024)}}, \emph{arXiv} \newline
{\small\color{metgray} 1960--2020, 2.3M scientists, 29.7M articles. $\cdot$ Data: MAG} \newline
Documents ``the death of Renaissance scientist''---proportion of generalists dropped from 55\% to 40\%. Since ``generalist teams are 31\% more likely to produce highly disruptive papers'' while ``specialist teams' likelihood of producing highly disruptive work declines by 22\%,``rising specialization contributes to declining disruption. & {\color{DeclineRed}\Large$\bm{\downarrow}$} \\
\addlinespace[1.2em]

\setcounter{inventry}{54}\refstepcounter{inventry}\label{inv:shinichi2024wp}\textcolor{metgray}{55} & \nociteinv{shinichi2024wp}\textbf{\href{https://doi.org/10.2139/ssrn.4706101}{\color{black}Shinichi \& Matsui (2024)}}, \emph{Working Paper} \newline
{\small\color{metgray} Visual Art (Ukiyo-e), 11,000 images. $\cdot$ Data: Ukiyo-e images} \newline
Analyzing visual content (VGGNet features) rather than citations, this study finds that ``the overall creativity of Ukiyo-e has been in decline since the late 1840s.'' The Disruption Index drops over time (Figs. 11--12), with early masters scoring systematically higher than later generations (Tables 4--5), indicating that measurement of declining disruption in creative domains is not an artifact of citation data.  & {\color{DeclineRed}\Large$\bm{\downarrow}$} \\
\addlinespace[1.2em]

\setcounter{inventry}{53}\refstepcounter{inventry}\label{inv:singh2024epjds}\textcolor{metgray}{54} & \nociteinv{singh2024epjds}\textbf{\href{https://doi.org/10.1140/epjds/s13688-024-00451-8}{\color{black}Singh et al. (2024)}}, \emph{\emph{EPJ Data Sci}} \newline
{\small\color{metgray} 1992--2018, 1.5M arXiv articles. $\cdot$ Data: arXiv} \newline
``Explorers'' more disruptive than ``exploiters'' (Fig. 5, p$<$0.001). Exploration peaks mid-career; ``senior scientists tend to exploit more'' (Fig. 4c). Combined with workforce aging, consistent with declining disruption. & {\color{DeclineRed}\Large$\bm{\downarrow}$} \\
\addlinespace[1.2em]

\setcounter{inventry}{52}\refstepcounter{inventry}\label{inv:sixt2024qss}\textcolor{metgray}{53} & \nociteinv{sixt2024qss}\textbf{\href{https://doi.org/10.1162/qss_a_00328}{\color{black}Sixt \& Pasin (2024)}}, \emph{\emph{Quant Sci Stud}} \newline
{\small\color{metgray} 1945--2010, 79M papers. $\cdot$ Data: Dimensions} \newline
Replicates decline using data from Dimensions database, across all fields of science, using CD index. Authors ``reproduce the decline of disruptive papers over time, which is one of the central results of Park et al. (2023)'' (Fig. 5). & {\color{DeclineRed}\Large$\bm{\downarrow}$} \\
\addlinespace[1.2em]

\setcounter{inventry}{51}\refstepcounter{inventry}\label{inv:tang2024ic}\textcolor{metgray}{52} & \nociteinv{tang2024ic}\textbf{\href{https://doi.org/10.1007/978-3-031-57860-1_27}{\color{black}Tang et al. (2024)}}, \emph{iConference} \newline
{\small\color{metgray} AI research, 1950--2019 (646,100 papers). $\cdot$ Data: MAG} \newline
The ``average disruptiveness percentile\ldots significantly and monotonically declines'' from 1950--2014, but ``since 2014\ldots [it] exhibit[s] startling upward trends'' as ``AI steps into the deep learning stage'' (Fig. 2b). & {\color{DeclineRed}\Large$\bm{\downarrow}$\hspace{-3pt}}{\color{IncreaseGreen}\Large$\bm{\uparrow}$} \\
\addlinespace[1.2em]

\setcounter{inventry}{50}\refstepcounter{inventry}\label{inv:tang2024joi}\textcolor{metgray}{51} & \nociteinv{tang2024joi}\textbf{\href{https://doi.org/10.1016/j.joi.2024.101570}{\color{black}Tang et al. (2024)}}, \emph{\emph{J Informetr}} \newline
{\small\color{metgray} AI researchers, 1961--2023, 665K papers. $\cdot$ Data: Scopus} \newline
Uses a novel, semantic-disruptiveness measure---``Disruptive Knowledge Content'' (Dkc) index---thereby avoiding concerns with citation data). Finds ``papers were becoming less disruptive over time,'' and that ``scholars who started their scientific research earlier tend to have higher disruptiveness'' (Fig. 6). & {\color{DeclineRed}\Large$\bm{\downarrow}$} \\
\addlinespace[1.2em]

\setcounter{inventry}{49}\refstepcounter{inventry}\label{inv:wang2024joi}\textcolor{metgray}{50} & \nociteinv{wang2024joi}\textbf{\href{https://doi.org/10.1016/j.joi.2024.101493}{\color{black}Wang et al. (2024)}}, \emph{\emph{J Informetr}} \newline
{\small\color{metgray} 3D printing patents, 1997--2019. $\cdot$ Data: PatSnap} \newline
Peak disruption in early years; ``disruption degrees of nodes in the main path show an overall decreasing trend'' (Fig. 10).  & {\color{DeclineRed}\Large$\bm{\downarrow}$} \\
\addlinespace[1.2em]

\setcounter{inventry}{48}\refstepcounter{inventry}\label{inv:wang2024eist}\textcolor{metgray}{49} & \nociteinv{wang2024eist}\textbf{\href{https://doi.org/10.1016/j.eist.2024.100836}{\color{black}Wang et al. (2024)}}, \emph{\emph{Environ Innov Soc Transit}} \newline
{\small\color{metgray} Climate mitigation, 1988--2017, 248K patents. $\cdot$ Data: USPTO} \newline
``CCMT patents parallel the pattern of utility patents, showing a marked decline [in disruption]'' (Fig. 5f).  & {\color{DeclineRed}\Large$\bm{\downarrow}$} \\
\addlinespace[1.2em]

\setcounter{inventry}{47}\refstepcounter{inventry}\label{inv:wu2024s}\textcolor{metgray}{48} & \nociteinv{wu2024s}\textbf{\href{https://doi.org/10.1007/s11192-024-05126-9}{\color{black}Wu et al. (2024)}}, \emph{\emph{Scientometrics}} \newline
{\small\color{metgray} Pharmaceuticals, 1940s--2018, 1.9M patents. $\cdot$ Data: MAG, PATSTAT} \newline
Shows technological disruptiveness declining (Fig. 3); authors note ``technological disruptiveness is diminishing over time.'' & {\color{DeclineRed}\Large$\bm{\downarrow}$} \\
\addlinespace[1.2em]

\setcounter{inventry}{46}\refstepcounter{inventry}\label{inv:yang2024bqss}\textcolor{metgray}{47} & \nociteinv{yang2024bqss}\textbf{\href{https://doi.org/10.1162/qss_a_00323}{\color{black}Yang \& Deng (2024)}}, \emph{\emph{Quant Sci Stud}} \newline
{\small\color{metgray} 1887--2010, 712 Nobel papers. $\cdot$ Data: MAG} \newline
Shows disruptiveness (CD index) declining (Fig. 2c); ``even for scientific breakthroughs, their disruptive potential diminishes over time.'' & {\color{DeclineRed}\Large$\bm{\downarrow}$} \\
\addlinespace[1.2em]

\setcounter{inventry}{45}\refstepcounter{inventry}\label{inv:yang2024joi}\textcolor{metgray}{46} & \nociteinv{yang2024joi}\textbf{\href{https://doi.org/10.1016/j.joi.2024.101588}{\color{black}Yang \& Wang (2024)}}, \emph{\emph{J Informetr}} \newline
{\small\color{metgray} Medicine, 1980--2021, 5M+ papers. $\cdot$ Data: MAG} \newline
Team freshness declining---from 0.41 to 0.13 (Fig. 3g); freshness of newcomers down 77\%. Freshness positively associated with disruption (Table 1, p$<$0.001). Declining freshness consistent with declining disruption. & {\color{DeclineRed}\Large$\bm{\downarrow}$} \\
\addlinespace[1.2em]

\setcounter{inventry}{44}\refstepcounter{inventry}\label{inv:yang2024wp}\textcolor{metgray}{45} & \nociteinv{yang2024wp}\textbf{\href{https://arxiv.org/abs/2311.16360v2}{\color{black}Yang \& Youn (2024)}}, \emph{arXiv} \newline
{\small\color{metgray} U.S. patents, 1981--2010. $\cdot$ Data: Google Patents, USPTO} \newline
Patent ``curvature'' rises over time (Fig. 2c); the authors find that ``inventions arising at high curvatures tend to be less disruptive'' (Table 1: $\beta$ = -0.002, p$<$0.001), implying declining disruption. & {\color{DeclineRed}\Large$\bm{\downarrow}$} \\
\addlinespace[1.2em]

\setcounter{inventry}{43}\refstepcounter{inventry}\label{inv:yang2024aqss}\textcolor{metgray}{44} & \nociteinv{yang2024aqss}\textbf{\href{https://doi.org/10.1162/qss_a_00335}{\color{black}Yang et al. (2024)}}, \emph{\emph{Quant Sci Stud}} \newline
{\small\color{metgray} STEM fields, 1980--2016, 3.2M papers. $\cdot$ Data: MAG} \newline
The probability of producing a disruptive paper (CD$>$0) declines steadily over time, falling from $\sim$25--30\% in 1980 to $\sim$15\% by 2016 for both male- and female-led teams (Fig. 3i). & {\color{DeclineRed}\Large$\bm{\downarrow}$} \\
\addlinespace[1.2em]

\setcounter{inventry}{42}\refstepcounter{inventry}\label{inv:yang2024s}\textcolor{metgray}{43} & \nociteinv{yang2024s}\textbf{\href{https://doi.org/10.1007/s11192-024-04987-4}{\color{black}Yang et al. (2024)}}, \emph{\emph{Scientometrics}} \newline
{\small\color{metgray} 1950--2017, 30M papers. $\cdot$ Data: SciSciNet} \newline
Teams grow older and less age-diverse over time (Fig. 1), and both shifts are associated with lower disruption (Table 4). & {\color{DeclineRed}\Large$\bm{\downarrow}$} \\
\addlinespace[1.2em]

\setcounter{inventry}{41}\refstepcounter{inventry}\label{inv:yang2023s}\textcolor{metgray}{42} & \nociteinv{yang2023s}\textbf{\href{https://doi.org/10.1007/s11192-023-04889-x}{\color{black}Yang et al. (2024)}}, \emph{\emph{Scientometrics}} \newline
{\small\color{metgray} Physics, 1980--2010, 300K papers. $\cdot$ Data: APS} \newline
Original DI shows declining trend (Fig. 7c). Rescaled (RDI) flat by design---removes temporal trends.  & {\color{DeclineRed}\Large$\bm{\downarrow}$} \\
\addlinespace[1.2em]

\setcounter{inventry}{40}\refstepcounter{inventry}\label{inv:yu2024s}\textcolor{metgray}{41} & \nociteinv{yu2024s}\textbf{\href{https://doi.org/10.1007/s11192-024-05184-z }{\color{black}Yu et al. (2024)}}, \emph{\emph{Scientometrics}} \newline
{\small\color{metgray} Quantum information, Nanomaterials, and Synthetic biology, 2000--2022, 1M papers. $\cdot$ Data: WoS} \newline
Rising interdisciplinary diversity is negatively associated with disruptiveness (Fig. 2), consistent with declining disruption overall, though field-level trends vary. Quantum information and Synthetic biology decline modestly (with a brief rebound around 2013); Nanomaterials spikes in 2002--2003 before stabilizing (Supp. Fig. 3). & {\color{MixedBlue}\Large$\bm{\updownarrow}$} \\
\addlinespace[1.2em]

\setcounter{inventry}{39}\refstepcounter{inventry}\label{inv:zhang2024joi}\textcolor{metgray}{40} & \nociteinv{zhang2024joi}\textbf{\href{https://doi.org/10.1016/j.joi.2024.101520 }{\color{black}Zhang et al. (2024)}}, \emph{\emph{J Informetr}} \newline
{\small\color{metgray} Physics papers, 1925--2016, 137K papers; robustness in Sociology and Psychology. $\cdot$ Data: APS} \newline
Year-disruption correlation: r=-0.102, p$<$0.001 (Table 3). Publication year remains negative predictor in regressions controlling for team characteristics (Tables 4--5). Sociology shows stronger decline ($\beta$=-0.191); Psychology mixed. & {\color{DeclineRed}\Large$\bm{\downarrow}$} \\
\addlinespace[1.2em]

\setcounter{inventry}{38}\refstepcounter{inventry}\label{inv:zhang2024wp}\textcolor{metgray}{39} & \nociteinv{zhang2024wp}\textbf{\href{https://arxiv.org/abs/2405.14543v2}{\color{black}Zhang et al. (2024)}}, \emph{arXiv} \newline
{\small\color{metgray} 1950s--2010s, 300K+ careers. $\cdot$ Data: SciSciNet} \newline
``Scientists nowadays make fewer disruptive contributions than those in the early years'' (Figs. 7c--d).  & {\color{DeclineRed}\Large$\bm{\downarrow}$} \\
\addlinespace[1.2em]

\setcounter{inventry}{37}\refstepcounter{inventry}\label{inv:zhang2024s}\textcolor{metgray}{38} & \nociteinv{zhang2024s}\textbf{\href{https://doi.org/10.1007/s11192-024-05134-9 }{\color{black}Zhang et al. (2024)}}, \emph{\emph{Scientometrics}} \newline
{\small\color{metgray} Chinese listed firm patents. $\cdot$ Data: Chinese patents} \newline
Compares D5 to Park et al. (2023): ``results indicate high consistency and correlation\ldots consistent [with an] overall downward trend'' (Fig. 3). Extends declining disruption evidence beyond U.S. to Chinese patent data. & {\color{DeclineRed}\Large$\bm{\downarrow}$} \\
\addlinespace[1.2em]

\\[1.5em]
\multicolumn{3}{c}{\large\textsc{2023} (17 studies)} \\[0.8em]
\setcounter{inventry}{36}\refstepcounter{inventry}\label{inv:bentley2023acs}\textcolor{metgray}{37} & \nociteinv{bentley2023acs}\textbf{\href{https://doi.org/10.1142/s0219525923500066}{\color{black}Bentley et al. (2023)}}, \emph{\emph{Adv Complex Syst}} \newline
{\small\color{metgray} Life sciences, 1930--2015. $\cdot$ Data: PubMed} \newline
Observes steep decline 1930--1950. Disruption ``close to zero for the 50 years after about 1970'' with ``modest but consistent growth\ldots  since 2000'' (Fig. 2). & {\color{DeclineRed}\Large$\bm{\downarrow}$\hspace{-3pt}}{\color{IncreaseGreen}\Large$\bm{\uparrow}$} \\
\addlinespace[1.2em]

\setcounter{inventry}{35}\refstepcounter{inventry}\label{inv:brauer2023wp}\textcolor{metgray}{36} & \nociteinv{brauer2023wp}\textbf{Br{\"a}uer (2023)}, \emph{Working Paper} \newline
{\small\color{metgray} Global patents, 1980--2010. $\cdot$ Data: PATSTAT} \newline
Using novel disruption metric (Citation Year Gap), tracks ``aggregate evolution of disruptive innovation'', showing CYG ``declines from -1 to almost -5'' across IPC classes, with the decline ``widespread'' (Figs. 2, 3). Concludes ``technology fields trend even further towards incrementalism.'' & {\color{DeclineRed}\Large$\bm{\downarrow}$} \\
\addlinespace[1.2em]

\setcounter{inventry}{34}\refstepcounter{inventry}\label{inv:gebhart2023wp}\textcolor{metgray}{35} & \nociteinv{gebhart2023wp}\textbf{\href{https://arxiv.org/abs/2308.16363v1}{\color{black}Gebhart \& Funk (2023)}}, \emph{arXiv} \newline
{\small\color{metgray} Physics, 1893--2019. $\cdot$ Data: APS} \newline
Measurement-robust decline. Using 7 mathematically distinct disruption metrics, the authors find a consistent downward trend on 6 of 7 measures (Fig. 8), ruling out metric-specific artifacts. & {\color{DeclineRed}\Large$\bm{\downarrow}$} \\
\addlinespace[1.2em]

\setcounter{inventry}{33}\refstepcounter{inventry}\label{inv:kaltenberg2023rp}\textcolor{metgray}{34} & \nociteinv{kaltenberg2023rp}\textbf{\href{https://doi.org/10.1016/j.respol.2022.104629}{\color{black}Kaltenberg et al. (2023)}}, \emph{\emph{Res Policy}} \newline
{\small\color{metgray} U.S. patents, 1976--2017, 3M patents, 1.2M inventors. $\cdot$ Data: USPTO} \newline
Inventor age negatively associated with disruption; ``disruptiveness of the average inventor's patents from their twenties is roughly three standard deviations greater than the average from their 60s and 70s'' (Fig. 9). Inventor workforce is also aging, consistent with declining disruptiveness. & {\color{DeclineRed}\Large$\bm{\downarrow}$} \\
\addlinespace[1.2em]

\setcounter{inventry}{32}\refstepcounter{inventry}\label{inv:leahey2023asr}\textcolor{metgray}{33} & \nociteinv{leahey2023asr}\textbf{\href{https://doi.org/10.1177/00031224231168074}{\color{black}Leahey et al. (2023)}}, \emph{\emph{Am Sociol Rev}} \newline
{\small\color{metgray} Citation Classics, 1931--1987, 2.5K papers. $\cdot$ Data: WoS} \newline
``Publication year is inversely related to the CD index, so recent papers tend to be less disruptive than older papers'' (Table 3). Robust to field, journal, author controls.  & {\color{DeclineRed}\Large$\bm{\downarrow}$} \\
\addlinespace[1.2em]

\setcounter{inventry}{31}\refstepcounter{inventry}\label{inv:lee2023wp}\textcolor{metgray}{32} & \nociteinv{lee2023wp}\textbf{\href{https://arxiv.org/abs/2308.14724v1}{\color{black}Lee \& Kim (2023)}}, \emph{arXiv} \newline
{\small\color{metgray} Marketing, 1978--2022, 8.6K articles. $\cdot$ Data: OpenAlex} \newline
Conceptual research is more disruptive (Fig. 2, p<.01 for JM, JMR) ``but is declining in publication quantity.'' Table 3 provides direct evidence of declining disruption---year fixed effects increasingly negative relative to 1978--1982 (p<.01). & {\color{DeclineRed}\Large$\bm{\downarrow}$} \\
\addlinespace[1.2em]

\setcounter{inventry}{30}\refstepcounter{inventry}\label{inv:lin2023n}\textcolor{metgray}{31} & \nociteinv{lin2023n}\textbf{\href{https://doi.org/10.1038/s41586-023-06767-1}{\color{black}Lin et al. (2023)}}, \emph{\emph{Nature}} \newline
{\small\color{metgray} 20M papers, 4M patents. $\cdot$ Data: MAG, USPTO} \newline
Reports declining disruptiveness; decline accelerates post-2000 (Ext. Data Tables 1--2). Remote collaboration rising; remote teams less likely to engage in ``early-stage, conceptual tasks,'' contributing to decline. Effect persists within same individuals, pronounced for junior researchers.  & {\color{DeclineRed}\Large$\bm{\downarrow}$} \\
\addlinespace[1.2em]

\setcounter{inventry}{29}\refstepcounter{inventry}\label{inv:park2023n}\textcolor{metgray}{30} & \nociteinv{park2023n}\textbf{\href{https://doi.org/10.1038/s41586-022-05543-x}{\color{black}Park et al. (2023)}}, \emph{\emph{Nature}} \newline
{\small\color{metgray} 45M papers (1945--2010), 3.9M patents (1976--2010). $\cdot$ Data: USPTO, WoS} \newline
Disruptiveness declines over time across papers and patents (Fig. 2a--b). ``Science and technology are becoming less disruptive.'' Demonstrates robustness on 4 additional datasets (JSTOR, APS, MAG, PubMed), Monte Carlo simulations, alternative metrics (ED Figs. 6--8). Provides corroborating evidence using text-based (non-citation) metrics. Highly disruptive works stable in absolute terms (Fig. 4). & {\color{DeclineRed}\Large$\bm{\downarrow}$} \\
\addlinespace[1.2em]

\setcounter{inventry}{28}\refstepcounter{inventry}\label{inv:sheng2023s}\textcolor{metgray}{29} & \nociteinv{sheng2023s}\textbf{\href{https://doi.org/10.1007/s11192-023-04751-0}{\color{black}Sheng et al. (2023)}}, \emph{\emph{Scientometrics}} \newline
{\small\color{metgray} Biomedicine, 2001--2010, 1.3M articles. $\cdot$ Data: PubMed} \newline
Reports that disruption decreased from 2001 to 2010 (Fig. 3). ``The maximum value shows a descending trend. Based on different citation windows, the interquartile range\ldots also decreases gradually, indicating that the discrete degree of the D index gradually decreases each year.'' & {\color{DeclineRed}\Large$\bm{\downarrow}$} \\
\addlinespace[1.2em]

\setcounter{inventry}{27}\refstepcounter{inventry}\label{inv:shu2023wp}\textcolor{metgray}{28} & \nociteinv{shu2023wp}\textbf{\href{https://arxiv.org/abs/2312.05927v1}{\color{black}Shu \& Pan (2023)}}, \emph{arXiv} \newline
{\small\color{metgray} 1960--2010, 25M papers. $\cdot$ Data: MAG} \newline
Using a novel text-based measure, the authors find that ``stylized papers [i.e., work that deviates from established paradigms] exhibit a higher likelihood of demonstrating disruption'' (Fig. 3), yet there is a ``concerning decline in stylized research'' over time (Fig. 1). Science now actively suppresses styled work; stylized papers face longer review times and receive fewer citations (Fig. 3), consistent with declining disruptiveness. & {\color{DeclineRed}\Large$\bm{\downarrow}$} \\
\addlinespace[1.2em]

\setcounter{inventry}{26}\refstepcounter{inventry}\label{inv:spinellis2023pone}\textcolor{metgray}{27} & \nociteinv{spinellis2023pone}\textbf{\href{https://doi.org/10.1371/journal.pone.0294946}{\color{black}Spinellis (2023)}}, \emph{\emph{PLoS ONE}} \newline
{\small\color{metgray} 1945--2016, 50.9M papers. $\cdot$ Data: Crossref} \newline
Using novel data, shows that CD index drops from 0.134 to 0.001 (Fig. 2). ``The fall in the CD index is in line with recently published research reporting that papers are becoming less disruptive over time.'' Reported correlation with Park et al.: $\rho$ = 0.95, p$<$0.001. & {\color{DeclineRed}\Large$\bm{\downarrow}$} \\
\addlinespace[1.2em]

\setcounter{inventry}{25}\refstepcounter{inventry}\label{inv:zutphen2023wp}\textcolor{metgray}{26} & \nociteinv{zutphen2023wp}\textbf{Van Zutphen (2023)}, \emph{Thesis} \newline
{\small\color{metgray} Marketing and AI, 1980--2020. $\cdot$ Data: WoS} \newline
Both fields exhibit periods of declining disruptiveness, but ``while AI papers still appear to have declined in disruptiveness in the most recent years, Marketing papers appear to have reversed their downward trend'' (Fig. 13). & {\color{MixedBlue}\Large$\bm{\updownarrow}$} \\
\addlinespace[1.2em]

\setcounter{inventry}{24}\refstepcounter{inventry}\label{inv:wang2023wp}\textcolor{metgray}{25} & \nociteinv{wang2023wp}\textbf{\href{https://doi.org/10.2139/ssrn.4923505}{\color{black}Wang (2023)}}, \emph{Working Paper} \newline
{\small\color{metgray} Biomedicine and life sciences, 1950--2021, 8.7M papers. $\cdot$ Data: PubMed} \newline
Using a citation-based disruption measure, the author finds that ``the disruptive nature of discoveries has declined'' over time, for both average and ``star'' (top 5\%) papers (Fig. 3). & {\color{DeclineRed}\Large$\bm{\downarrow}$} \\
\addlinespace[1.2em]

\setcounter{inventry}{23}\refstepcounter{inventry}\label{inv:wu2023joi}\textcolor{metgray}{24} & \nociteinv{wu2023joi}\textbf{\href{https://doi.org/10.1016/j.joi.2023.101436}{\color{black}Wu et al. (2023)}}, \emph{\emph{J Informetr}} \newline
{\small\color{metgray} Economics (ML vs. non-ML), 2001--2020. $\cdot$ Data: MAG} \newline
Using a standard citation-based disruption index, ML-related economics papers are significantly more disruptive than non-ML work, with the gap widening after 2010 (Fig. 7). This pattern suggests that disruption can increase under methodological regime shifts, providing evidence against claims that observed declines are purely mechanical. & {\color{IncreaseGreen}\Large$\bm{\uparrow}$} \\
\addlinespace[1.2em]

\setcounter{inventry}{22}\refstepcounter{inventry}\label{inv:yang2023ipm}\textcolor{metgray}{23} & \nociteinv{yang2023ipm}\textbf{\href{https://doi.org/10.1016/j.ipm.2023.103420}{\color{black}Yang et al. (2023)}}, \emph{\emph{Inf Process Manag}} \newline
{\small\color{metgray} Physics, 1893--2010, 463K papers. $\cdot$ Data: APS} \newline
Uses Disruptive Citation measure (resistant to reference-count manipulation). Shows decline from $\sim$1960s onward (Fig. B1f). & {\color{DeclineRed}\Large$\bm{\downarrow}$} \\
\addlinespace[1.2em]

\setcounter{inventry}{21}\refstepcounter{inventry}\label{inv:yang2023issip}\textcolor{metgray}{22} & \nociteinv{yang2023issip}\textbf{Yang et al. (2023)}, \emph{ISSI} \newline
{\small\color{metgray} Information science, 1978--1989. $\cdot$ Data: WoS} \newline
Using Global Disruption Index (GDI), designed to address limitations of previous measures, reports ``average disruptiveness has declined\ldots papers are less likely to be disruptive'' (Fig. 5). Further, also observe that ``early-published papers tend to be more disruptive as they are more likely to challenge existing scientific paradigms.'' & {\color{DeclineRed}\Large$\bm{\downarrow}$} \\
\addlinespace[1.2em]

\setcounter{inventry}{20}\refstepcounter{inventry}\label{inv:zeng2023wp}\textcolor{metgray}{21} & \nociteinv{zeng2023wp}\textbf{\href{https://arxiv.org/abs/2305.03589v1}{\color{black}Zeng et al. (2023)}}, \emph{arXiv} \newline
{\small\color{metgray} Physics and five other disciplines. $\cdot$ Data: AMiner, APS, MAG} \newline
Documents sharp decline in disruptiveness over time, especially among highly cited papers---``disruptive papers in science are losing impact.'' Fraction with positive disruption (D5$>$0) also declines over time. & {\color{DeclineRed}\Large$\bm{\downarrow}$} \\
\addlinespace[1.2em]

\\[1.5em]
\multicolumn{3}{c}{\large\textsc{2022} (8 studies)} \\[0.8em]
\setcounter{inventry}{19}\refstepcounter{inventry}\label{inv:abu-omar2022cr}\textcolor{metgray}{20} & \nociteinv{abu-omar2022cr}\textbf{\href{https://doi.org/10.1016/j.crad.2022.07.003}{\color{black}Abu-Omar et al. (2022)}}, \emph{\emph{Clin Radiol}} \newline
{\small\color{metgray} Radiology, 1954--2014. $\cdot$ Data: PubMed} \newline
Downward trend (Fig. 3). ``Most papers published showed negligible or negative disruption scores.'' & {\color{DeclineRed}\Large$\bm{\downarrow}$} \\
\addlinespace[1.2em]

\setcounter{inventry}{18}\refstepcounter{inventry}\label{inv:becerra2022dcr}\textcolor{metgray}{19} & \nociteinv{becerra2022dcr}\textbf{\href{https://doi.org/10.1097/dcr.0000000000002118}{\color{black}Becerra et al. (2022)}}, \emph{\emph{Dis Colon Rectum}} \newline
{\small\color{metgray} Colorectal surgery, 1954--2014. $\cdot$ Data: PubMed} \newline
``The decade with the most articles was 1980 to 1989 and the decade with the least\ldots was 2000 to 2009'' (Fig. 4).  & {\color{DeclineRed}\Large$\bm{\downarrow}$} \\
\addlinespace[1.2em]

\setcounter{inventry}{17}\refstepcounter{inventry}\label{inv:bessen2022disruption}\textcolor{metgray}{18} & \nociteinv{bessen2022disruption}\textbf{\href{https://doi.org/10.2307/j.ctv2jn91q7.6}{\color{black}Bessen (2022)}}, \emph{Yale Univ Press} \newline
{\small\color{metgray} U.S. industries, 1970s--2010s. $\cdot$ Data: Compustat, NETS} \newline
Documents ``declining industrial disruption'' using non-citation-based measure. ``The likelihood that a top firm in any industry will be displaced by a rival is less than half of what it was in the late 1990s,'' as ``the rate of disruption sharply reversed trend around 2000'' (Fig. 1). & {\color{DeclineRed}\Large$\bm{\downarrow}$} \\
\addlinespace[1.2em]

\setcounter{inventry}{16}\refstepcounter{inventry}\label{inv:chen2022wp}\textcolor{metgray}{17} & \nociteinv{chen2022wp}\textbf{Chen et al. (2022)}, \emph{Working Paper} \newline
{\small\color{metgray} Scientific teams, 1961--2020. $\cdot$ Data: MAG} \newline
The authors show that ``the percentage of fundamental or disruptive scientific discoveries has declined'' in tandem with the rise of distributed teams (Fig. 1). While improved ICT mitigates this effect after 2010, the long-run decline in disruption persists. & {\color{DeclineRed}\Large$\bm{\downarrow}$} \\
\addlinespace[1.2em]

\setcounter{inventry}{15}\refstepcounter{inventry}\label{inv:cui2022wp}\textcolor{metgray}{16} & \nociteinv{cui2022wp}\textbf{\href{https://arxiv.org/abs/2202.04044v1}{\color{black}Cui et al. (2022)}}, \emph{arXiv} \newline
{\small\color{metgray} 1960--2015, 244M scholars. $\cdot$ Data: MAG, WoS} \newline
Shows the 1990s cohort had lower disruption than the 1960s at the same career age (Fig. S7B). ``As scientists age\ldots research is less likely to disrupt.'' Also shows the workforce is aging, further corroborating declining disruptiveness (Fig. S5). & {\color{DeclineRed}\Large$\bm{\downarrow}$} \\
\addlinespace[1.2em]

\setcounter{inventry}{14}\refstepcounter{inventry}\label{inv:peters2022wp}\textcolor{metgray}{15} & \nociteinv{peters2022wp}\textbf{Peters \& Walsh (2022)}, \emph{Working Paper} \newline
{\small\color{metgray} 1980--2012. $\cdot$ Data: Census LBD} \newline
Provides evidence of declining disruption using economic rather than bibliometric data. ``The decline in population growth has contributed to\ldots a decline in firm entry and in the rate of `creative destruction''' (Fig. 1, right panel). This suggests the trend is visible in market dynamics independent of citation metrics. & {\color{DeclineRed}\Large$\bm{\downarrow}$} \\
\addlinespace[1.2em]

\setcounter{inventry}{13}\refstepcounter{inventry}\label{inv:singh2022pone}\textcolor{metgray}{14} & \nociteinv{singh2022pone}\textbf{\href{https://doi.org/10.1371/journal.pone.0270131}{\color{black}Singh et al. (2022)}}, \emph{\emph{PLoS ONE}} \newline
{\small\color{metgray} 1986--2018, 175 fields, 1.5M arXiv articles. $\cdot$ Data: arXiv} \newline
Disruption highest in early ``creation'' stages, declines as fields mature (Fig. 4). ``Disruptiveness [is] fading.''  & {\color{DeclineRed}\Large$\bm{\downarrow}$} \\
\addlinespace[1.2em]

\setcounter{inventry}{12}\refstepcounter{inventry}\label{inv:you2022joi}\textcolor{metgray}{13} & \nociteinv{you2022joi}\textbf{\href{https://doi.org/10.1016/j.joi.2022.101294}{\color{black}You et al. (2022)}}, \emph{\emph{J Informetr}} \newline
{\small\color{metgray} 1996--2018, 48.6M papers. $\cdot$ Data: Scopus} \newline
Shows ``market share of questionable [predatory] publications has steadily increased'' (Fig. S1) and that ``questionable publications were less disruptive\ldots than their counterparts'' ($D \sim 0.05$ vs $\sim 0.14$, Fig. 4C), providing compositional evidence for declining disruption. & {\color{DeclineRed}\Large$\bm{\downarrow}$} \\
\addlinespace[1.2em]

\\[1.5em]
\multicolumn{3}{c}{\large\textsc{2021} (7 studies)} \\[0.8em]
\setcounter{inventry}{11}\refstepcounter{inventry}\label{inv:bornmann2021joi}\textcolor{metgray}{12} & \nociteinv{bornmann2021joi}\textbf{\href{https://doi.org/10.1016/j.joi.2021.101159}{\color{black}Bornmann \& Tekles (2021)}}, \emph{\emph{J Informetr}} \newline
{\small\color{metgray} Physics, 1980--2002. $\cdot$ Data: WoS} \newline
Validates multiple disruption indices against expert milestones. Notes ``a time dependency of disruption, with high disruptive values for older milestone papers'' (Fig. 3).  & {\color{DeclineRed}\Large$\bm{\downarrow}$} \\
\addlinespace[1.2em]

\setcounter{inventry}{10}\refstepcounter{inventry}\label{inv:chu2021pnas}\textcolor{metgray}{11} & \nociteinv{chu2021pnas}\textbf{\href{https://doi.org/10.1073/pnas.2021636118}{\color{black}Chu \& Evans (2021)}}, \emph{\emph{PNAS}} \newline
{\small\color{metgray} 1960--2014, 90M papers. $\cdot$ Data: WoS} \newline
Disruptive share drops from $\sim$49\% to 13\% as fields grow (Fig. 4). Concludes, ``newly published papers become unlikely to disrupt existing work.'' & {\color{DeclineRed}\Large$\bm{\downarrow}$} \\
\addlinespace[1.2em]

\setcounter{inventry}{9}\refstepcounter{inventry}\label{inv:grunvald2021apjcp}\textcolor{metgray}{10} & \nociteinv{grunvald2021apjcp}\textbf{\href{https://doi.org/10.31557/apjcp.2021.22.8.2385}{\color{black}Grunvald et al. (2021)}}, \emph{\emph{Asian Pac J Cancer Prev}} \newline
{\small\color{metgray} Breast cancer, 1954--2014. $\cdot$ Data: PubMed} \newline
Reports that the median publication years of the top 100 is 1977. ``Disruptive studies skewed towards being older'' (Fig. 3).  & {\color{DeclineRed}\Large$\bm{\downarrow}$} \\
\addlinespace[1.2em]

\setcounter{inventry}{8}\refstepcounter{inventry}\label{inv:hansdorfer2021prsgo}\textcolor{metgray}{9} & \nociteinv{hansdorfer2021prsgo}\textbf{\href{https://doi.org/10.1097/gox.0000000000003446}{\color{black}Hansdorfer et al. (2021)}}, \emph{\emph{Plast Reconstr Surg Glob Open}} \newline
{\small\color{metgray} Plastic surgery, 1954--2014. $\cdot$ Data: PubMed} \newline
The authors find the 1980s had the most disruptive papers, while the 2000s had the least (Fig. 4).  & {\color{DeclineRed}\Large$\bm{\downarrow}$} \\
\addlinespace[1.2em]

\setcounter{inventry}{7}\refstepcounter{inventry}\label{inv:horen2021jcs}\textcolor{metgray}{8} & \nociteinv{horen2021jcs}\textbf{\href{https://doi.org/10.1097/scs.0000000000007804}{\color{black}Horen et al. (2021)}}, \emph{\emph{J Craniofac Surg}} \newline
{\small\color{metgray} Craniofacial surgery, 1954--2014. $\cdot$ Data: PubMed} \newline
Finds among top disruptive publications, decade with the most was 1980s, 2000s was least (Fig. 4). 1960s highest average disruptiveness; 2000s the lowest.  & {\color{DeclineRed}\Large$\bm{\downarrow}$} \\
\addlinespace[1.2em]

\setcounter{inventry}{6}\refstepcounter{inventry}\label{inv:sullivan2021jps}\textcolor{metgray}{7} & \nociteinv{sullivan2021jps}\textbf{\href{https://doi.org/10.1016/j.jpedsurg.2021.02.002}{\color{black}Sullivan et al. (2021)}}, \emph{\emph{J Pediatr Surg}} \newline
{\small\color{metgray} Pediatric surgery, 1954--2014. $\cdot$ Data: PubMed} \newline
Among highly disruptive articles, ``1970s and 1980s were the most represented decades'' with decline thereafter (Fig. 3).  & {\color{DeclineRed}\Large$\bm{\downarrow}$} \\
\addlinespace[1.2em]

\setcounter{inventry}{5}\refstepcounter{inventry}\label{inv:zeng2021nhb}\textcolor{metgray}{6} & \nociteinv{zeng2021nhb}\textbf{\href{https://doi.org/10.1038/s41562-021-01084-x }{\color{black}Zeng et al. (2021)}}, \emph{\emph{Nat Hum Behav}} \newline
{\small\color{metgray} Physics, 1893--2010, 480K+ papers. $\cdot$ Data: APS} \newline
Studies disruption and multidisciplinary impact, finding that ``both indexes decrease with time.'' (Supp. Fig. 6). & {\color{DeclineRed}\Large$\bm{\downarrow}$} \\
\addlinespace[1.2em]

\\[1.5em]
\multicolumn{3}{c}{\large\textsc{2020} (2 studies)} \\[0.8em]
\setcounter{inventry}{4}\refstepcounter{inventry}\label{inv:bessen2020wp}\textcolor{metgray}{5} & \nociteinv{bessen2020wp}\textbf{\href{https://doi.org/10.2139/ssrn.3682745}{\color{black}Bessen et al. (2020)}}, \emph{Working Paper} \newline
{\small\color{metgray} U.S. industries, 1976--2017. $\cdot$ Data: Compustat, NETS} \newline
Non-citation measure shows ``declining industrial disruption.'' Firm displacement hazards ``rose from 1970 until the late 1990s'' then ``sharply reversed'' around 2000, with ``substantial declines in the displacement rate'' (Fig. 1). & {\color{DeclineRed}\Large$\bm{\downarrow}$} \\
\addlinespace[1.2em]

\setcounter{inventry}{3}\refstepcounter{inventry}\label{inv:falcao2020pismirc}\textcolor{metgray}{4} & \nociteinv{falcao2020pismirc}\textbf{\href{https://doi.org/10.5281/zenodo.4245355}{\color{black}Falc{\~a}o et al. (2020)}}, \emph{ISMIR} \newline
{\small\color{metgray} Forró music, 1945--2016, 27,352 songs. $\cdot$ Data: Forró music} \newline
Measuring disruption from raw audio features, the study shows that ``the higher disruptions of Forró are mostly concentrated on its first years\ldots between 1960 and 1970,'' with consolidation thereafter (Fig. 8). Because this approach avoids citation data entirely, the finding suggests that declining disruption in creative domains is not driven by citation-based artifacts. & {\color{DeclineRed}\Large$\bm{\downarrow}$} \\
\addlinespace[1.2em]

\\[1.5em]
\multicolumn{3}{c}{\large\textsc{2019} (2 studies)} \\[0.8em]
\setcounter{inventry}{2}\refstepcounter{inventry}\label{inv:bowen2019wp}\textcolor{metgray}{3} & \nociteinv{bowen2019wp}\textbf{\href{https://doi.org/10.2139/ssrn.3478503}{\color{black}Bowen et al. (2019)}}, \emph{Working Paper} \newline
{\small\color{metgray} U.S. patents, 1930--2010. $\cdot$ Data: USPTO} \newline
Text-based disruption. The ``average disruptive potential of U.S. patents has markedly declined since 1950,'' falling to ``roughly one quarter'' of its peak by 2010, even after adjusting for technology class and location (Fig. 2); the decline in technological disruptiveness indicates a widespread deceleration in vocabulary usage growth rates.'' & {\color{DeclineRed}\Large$\bm{\downarrow}$} \\
\addlinespace[1.2em]

\setcounter{inventry}{1}\refstepcounter{inventry}\label{inv:wu2019n}\textcolor{metgray}{2} & \nociteinv{wu2019n}\textbf{\href{https://doi.org/10.1038/s41586-019-0941-9}{\color{black}Wu et al. (2019)}}, \emph{\emph{Nature}} \newline
{\small\color{metgray} 1954--2014, 65M+ papers, patents, and software. $\cdot$ Data: GitHub, USPTO, WoS} \newline
Decline in disruption observed across all team sizes over time. ``Earlier cohorts are more disruptive than later cohorts'' (ED Fig. 3c). Small teams are more disruptive but have declined in prevalence (ED Fig. 8).  & {\color{DeclineRed}\Large$\bm{\downarrow}$} \\
\addlinespace[1.2em]

\\[1.5em]
\multicolumn{3}{c}{\large\textsc{2017} (1 studies)} \\[0.8em]
\setcounter{inventry}{0}\refstepcounter{inventry}\label{inv:funk2017ms}\textcolor{metgray}{1} & \nociteinv{funk2017ms}\textbf{\href{https://doi.org/10.1287/mnsc.2015.2366}{\color{black}Funk \& Owen-Smith (2017)}}, \emph{\emph{Manage Sci}} \newline
{\small\color{metgray} U.S. patents, 1977--2005, 2.9M patents. $\cdot$ Data: USPTO} \newline
Reports a decline in disruptiveness using the CD index, noting ``the average patent of today is less destabilizing than those of the late 1970s and early 1980s.'' Also finds ``significant (all p$<$0.001) negative correlations across categories between application and grant year and the CD5 index.'' & {\color{DeclineRed}\Large$\bm{\downarrow}$}

\end{longtable}